\documentclass[12pt]{article}
\usepackage{amsmath,amssymb,cite}
\usepackage[dvips]{graphicx}
\usepackage[dvips]{color}
\numberwithin{equation}{section}
\newcommand{\Cb}{{\mathbb C}}
\newcommand{\Cx}{{\mathbb C}^{\times}}
\newcommand{\Zb}{{\mathbb Z}}
\newcommand{\Rb}{{\mathbb R}}
\newcommand{\Ncal}{{\cal N}}
\newcommand{\del}{\partial}

\newcommand{\deltam}[2]{\delta^{{\rm mod}\;#2}_{#1}}
\newcommand{\ch}{\hat{c}}

\newcommand{\Th}{\Theta}
\newcommand{\e}[1]{\,{\bf e}\!\left[#1\right]}

\newcommand{\lcm}{{\rm lcm}}

\newcommand{\taub}{\bar{\tau}}

\newcommand{\nn}{\nonumber}

\DeclareMathOperator*{\Tr}{{\rm Tr}}
\newcommand{\vb}{{\vec{\beta}}}
\newcommand{\ve}{\varepsilon}
\newcommand{\rra}{\rangle\rangle}
\newcommand{\lla}{\langle\langle}
\newcommand{\ra}{\rangle}

\newcommand{\taut}{\tilde{\tau}}
\newcommand{\qt}{\tilde{q}}
\newcommand{\nom}{\nonumber}
\newcommand{\non}{\nonumber\\}
\newcommand{\dis}{\displaystyle}
\newcommand{\ba}{\begin{eqnarray}}
\newcommand{\ea}{\end{eqnarray}}
\newcommand{\cha}{\mbox{ch}}
\newcommand{\td}{\mbox{Td}}
\newcommand{\co}{{\cal O}}
\newcommand{\gam}{\Gamma}
\newcommand{\gyoretu}[1]{\begin{matrix}#1\end{matrix}}
\newcommand{\mfd}{{\cal M}}

\renewcommand{\a}{{\sf a}}
\newcommand{\0}{{\sf 0}}
\newcommand{\K}{K'}
\newcommand{\IL}{{I_{\blacklozenge}}}
\begin{document}
\baselineskip 3.5ex

%
%
\newcommand{\KUCPlogo}{\hbox{\lower 1.4ex\hbox{
\Huge\boldmath $\cal K$}
\kern -1.15em {\sffamily \bfseries\large\ UCP}}
\kern -4.7em \raise 0.3em\hbox{\lower 1.4ex\hbox{\color{cyan}
\Huge\boldmath $\cal K$}
\kern -1.15em {\color{magenta}\sffamily \bfseries\large\ UCP}
\put(-25,-7){\tiny\it preprint}
}}

\newcommand{\preprintnumber}{KUCP-0172 \\
 {\tt hep-th/0011091}}

\thispagestyle{empty}
\hbox{\raise 0.3ex\KUCPlogo\hspace{11cm}
\parbox[b]{4cm}{\preprintnumber}}
\noindent\hbox{\raise 1.5 ex \hbox{\rule{\textwidth}{0.3pt}}}

\vspace{20mm}\noindent
{\sffamily \bfseries \LARGE \boldmath
D-branes on a Noncompact Singular Calabi-Yau Manifold 
}

\vspace{0.5cm}\noindent
\rule{\textwidth}{1pt}

\vspace{1cm}

\noindent
{\sffamily \bfseries \large Katsuyuki Sugiyama}

\vspace{5mm}
\noindent
\hspace{0.7cm} \parbox{142mm}{\it
Department of Fundamental Sciences,
Faculty of Integrated Human Studies,
Kyoto University, Yoshida-Nihonmatsu-cho,
Sakyo-ku, Kyoto 606-8501, Japan.

\vspace*{3mm}
E-mail: {\tt sugiyama@phys.h.kyoto-u.ac.jp}
}

\vspace{5mm}

\noindent
{\sffamily \bfseries \large Satoshi Yamaguchi}

\vspace{5mm}
\noindent
\hspace{0.7cm} \parbox{142mm}{\it
Graduate School of Human and Environmental Studies,
Kyoto University, Yoshida-Nihonmatsu-cho,
Sakyo-ku, Kyoto 606-8501, Japan.

\vspace*{3mm}
E-mail: {\tt yamaguch@phys.h.kyoto-u.ac.jp}
}

\vspace{20mm}

\noindent {\sc Abstract:\ \ } We investigate D-branes on a noncompact
singular Calabi-Yau manifold by using the boundary CFT description, and
calculate the open string Witten indices between the boundary states.
The B-type D-branes turn out to be characterized by the properties of a
compact positively curved manifold.  We give geometric interpretations
to these boundary states in terms of coherent sheaves of the manifold.

%

\newpage
%
%

\section{Introduction}
D-branes are the key objects by which we can study the nature of various
string theories. The Cardy's boundary CFT method \cite{Car89} is a very
powerful tool to investigate the D-branes in curved spaces.

Recently, there has been great progress\cite{BDLR9906,Dou9910} in the study
of properties of charges, boundary states\cite{Ish89,Car89} based on Gepner
models\cite{Gep87,Gep88}.  For a large class of Calabi-Yau manifolds,
associated boundary states have been constructed and susy cycles have
been investigated based on these states in
CFTs\cite{OOY,RS9712,RS9811,GS9808,GS9902},
\cite{BDLR9906,Dou9910,BPPZ9908,DR9910,KLLW9912,%
Sch9912,NN0001,BS0001,FSW0003}.  These references consider D-branes as the
``rational boundary states'' , that is, the Cardy states realized as
some linear combinations of tensor products of Ishibashi states of each
minimal model.  There appear many consistency checks about their
charges, the intersection form of the homology cycles.  However these
analyses are restricted to compact Calabi-Yau cases.

In \cite{GKP9907}, Giveon et al. proposed that a string theory on a
noncompact singular Calabi-Yau manifold $X$ represented by a
hypersurface $ F(z_1,\dots,z_{n+1})=0 \text{ in }\Cb^{n+1}$ can be
described by a CFT as $ \Rb_{\phi}\times S^1 \times LG(W=F)$.  Here
$\Rb_{\phi}$ is a linear dilaton and $LG(W=F)$ is the two dimensional
$\Ncal=2$ Landau-Ginzburg theory with the superpotential $F$.

Also, in \cite{ES0002,Miz0003,Yam0007,Miz0009,NN0010}, the modular
invariant partition functions can be constructed in the cases that
$LG(W=F)$'s are minimal models or direct products of minimal models, and
it is shown that associated string theories on these singular spaces can
exist consistently.  They are extensions of Gepner models from compact
manifolds to noncompact varieties with singularities.

The aim of this paper is to develop a method to construct boundary
states of noncompact singular Calabi-Yau manifolds and to investigate
their properties in order to understand structures of moduli spaces in
the open string channel.  In this paper, we explore the ``rational''
D-branes on a noncompact singular Calabi-Yau manifold by applying the
Cardy's method to the Gepner-like description of the noncompact singular
Calabi-Yau manifold.

As a result, the open string Witten index turns out to be factorized
into $\theta_1(\tau)(=0)$ factor and a nontrivial one, in the same way
as the closed string Witten index\cite{Yam0007} does.  We investigate
the nontrivial factor in the open string Witten index and show that they
coincide with some pairings of bundles in the manifold $X/\Cx$.

To confirm the validity of this claim, we calculate the relative Euler
characteristics of the bundles in $X/\Cx$ by applying geometrical
methods to a special case, where $X/\Cx$ is written as a Fermat type
hypersurface $z_1^N+\dots+z_{r}^N=0$ in $\Cb P^{r-1}$.  We compare it
with the result in the CFT calculations.

The paper is organized as follows. In section 2, we review the
Gepner-like description of noncompact singular Calabi-Yau manifolds and
fix our convention.  In section 3, we construct the boundary states of
the noncompact manifold in the Gepner-like description and study
intersection number between these states.  This intersection pairing is
understood as a combination of characteristic classes with a pair of
boundary bundles and we compare the result in the CFT with the
geometrical ones.  We give geometrical interpretations to the boundary
states in terms of coherent sheaves in section 4.  Charges of D-branes
wrapped on the susy cycles are realized as characteristic classes of the
sheaves.  Section 5 is devoted to conclusions and discussions.  In
appendix A, we summarize several useful properties of theta functions.
Periods near the orbifold point are collected in appendix B. Also
formulas of periods in the large volume region are shown concretely in
appendix C.

\section{The closed string theory on a noncompact singular Calabi-Yau
manifold}\label{ClosedString}

In this section, we summarize the Gepner-like description of the closed
string theory on a noncompact singular Calabi-Yau $n$-fold $X$. We
mainly use the same conventions as in the paper\cite{Yam0007}.

We assume that the $X$ is realized as a zero locus $F(z_1,\dots,z_{n+1})=0$
 in $\Cb^{n+1}$, where $F(z)$ is a quasi-homogeneous polynomial,
 i.e. it satisfies the relation
\begin{eqnarray}
F(\lambda^{r_1}z_1,\dots,\lambda^{r_{n+1}}z_{n+1}) =\lambda
F(z_1,\dots,z_{n+1})\qquad \mbox{for} \,\,\,{}^{\exists}r_j\in
\Rb,{}^\forall\lambda\in\Cx\,.\nom
\end{eqnarray}
It is proposed in \cite{GKP9907} that the string theory on $X$ is
described by a model
\begin{eqnarray*}
 && \Rb_{\phi}\times S^1 \times LG(W=F),
\end{eqnarray*}
where $\Rb_{\phi}$ is a real line with a linear dilaton background and
$LG(W=F)$ is a scale invariant theory realized in the IR limit of the
Landau-Ginzburg theory with the superpotential $F$. In this paper we
consider the case where $LG(W=F)$ is described as a direct product of
A-type minimal models, namely, the polynomial $F$ is written as a linear
combination of constituent minimal models
\begin{eqnarray}
 && F(z_1,\dots,z_{n+1})=z_1^{N_1}+\dots+z_r^{N_r}+z_{r+1}^2+
\dots+z_{n+1}^2.\label{QuasiHomogeneous}
\end{eqnarray}
Here the Landau-Ginzburg theory is composed of $r$ minimal models and
the level of the $j$-th minimal model is $N_j-2$.

The remaining parts $\Rb_{\phi}\times S^1$ also have worldsheet
$\Ncal=2$ superconformal symmetry. We denote the bosonic coordinates of
$\Rb_{\phi}$ and $S^1$ by $\phi$ and $Y$, respectively, and the
fermionic counterparts of $\Rb_{\phi}$ and $S^1$ by the free fermions
$\psi^{\pm}$.  Then the $\Ncal=2$ superconformal currents are expressed
as
\begin{eqnarray}
 && T=-\frac12(\del Y)^2-\frac12(\del\phi)^2-\frac Q2 \del^2 \phi
-\frac12(\psi^{+}\del\psi^{-}-\del \psi^{+}\psi^{-}),\nn\\
 &&G^{\pm}=-\frac{1}{\sqrt 2}\psi^{\pm}(i\del Y\pm\del\phi)\mp
\frac{Q}{\sqrt2}\del\psi^{\pm},\nn\\
 && J=\psi^{+}\psi^{-}- Qi\del Y. \label{LiouvilleSCA}
\end{eqnarray}
The Liouville field $\phi$ has a background charge $Q$ and the
associated central charge of this algebra is given as
$\ch\;(:=c/3)=1+Q^2$.  Since we are considering the string theory on a
Calabi-Yau $n$-fold, the central charge should satisfy the consistency
condition $\ch=n$. When we take into account of the background charge
$Q$ of $\Rb_{\phi}$, the condition on the central charge is represented
as
\begin{eqnarray*}
 1+Q^2+\sum_{j=1}^{r}\frac{N_j-2}{N_j}=n.
\end{eqnarray*}
It means that the background charge $Q$ is determined from the
parameters of the minimal models $r$, $N_j$ through a relation
\begin{eqnarray*}
 &&Q^2=n-1-r+\sum_{j=1}^{r}\frac{2}{N_j}.
\end{eqnarray*}
For simplicity, we concentrate on the case that $n-1-r\equiv 0 \mod 2$
in this paper. Then $KQ^2$ is an even integer with $K=\lcm(N_j)$ and we
can define an integer $J$ as
\begin{eqnarray*}
 J:=\frac{KQ^2}{2}\in{\Zb}.
\end{eqnarray*}

Next, let us construct the modular invariant partition function of this
model and determine the spectrum.  As a consistency condition, we have
to pick up only the states with integral $U(1)$ charges since we want a
Calabi-Yau CFT.  First the linear dilaton $\phi$ does not have any
$U(1)$ charge and we can independently consider the partition function
$Z_{L}$ of $\phi$ \cite{ES0002}
\begin{eqnarray*}
 Z_{L}(\tau,\taub) = \frac{1}{\sqrt{\tau_2}|\eta(\tau)|^2},
\end{eqnarray*}
where $\eta(\tau)$ is the Dedekind eta function.

The other parts, $Y,\psi^{\pm}$ and fields in the minimal models carry
$U(1)$ charges, and we must apply the GSO projection to them.  Let us
consider first the case of $Y$.  If we consider the Verma module with
primary field $e^{ipY}$ for a real number $p$, the character of this
Verma module is $q^{p^2/2}/\eta(\tau)$, where $q=\exp (2\pi i
\tau)$. One can see the $U(1)$ charge of the states in this Verma module
is $pQ$.  We should take the case that $KpQ$ is an integer, since the
$U(1)$ charge of the other part is $(\text{integer})/K$ and the total
$U(1)$ charge should be an integer.  If we write $KpQ=2KJu+m_0$ with two
integers $u,m_0$ and sum up the character for all $u \in \Zb$, then we
obtain the relation
\begin{eqnarray*}
 \sum_{u\in \Zb} \frac{q^{\frac12 p^2}}{\eta(\tau)}
=\frac{\Th_{m_0,KJ}(\tau)}{\eta(\tau)}.
\end{eqnarray*}
Note that the $U(1)$ charges of the states included in these Verma
module are $m_0/K \mod 2J$.

Next we shall consider fermionic parts.  The Verma modules of two
fermions $\psi^{\pm}$ are characterized by an integer $s_0=0,1,2,3$. The
states with $s_0=0,2$ belong to the NS sector, and ones with $s_0=1,3$
are states in the R sector.  The characters of these Verma modules are
expressed by using theta functions $\Th_{s_0,2}(\tau)/\eta(\tau)$.

Last we study the Verma modules of a minimal model.  An arbitrary Verma
module in this model is labelled by three integers $(\ell,m,s)$.  They
take their values in the following ranges
\begin{eqnarray}
 && \ell=0,1,\dots,N-2,\nn\\
 && m=0,1,\dots,2N-1,\nn\\
 && s=0,1,2,3,\nn\\
 && \ell+m+s\equiv 0 \mod 2. \label{EvenCondition}
\end{eqnarray}
The states with $s_0=0,2$ belong to the NS sector, and ones with
$s_0=1,3$ are states in the R-sector. From now on, we denote the
character of this Verma module as $\chi_{m}^{\ell,s}(\tau)$

In order to consider the whole Verma module of a set of minimal models,
we introduce the following vector notations
\begin{align*}
  \vec \ell=(\ell_1,\dots,\ell_r),\quad
  \vec m =(m_0,m_1,\dots,m_r), \quad
  \vec s =(s_0,s_1,\dots,s_r),\quad
   \a=(\vec \ell, \vec m, \vec s)
\end{align*}
where $ (\ell_j,m_j,s_j),\ j=1,\dots,r$ is a set of indices of the Verma
module in the $j$-th minimal model and $m_0$ and $s_0$ were defined
above.  By collecting contributions of constituent minimal models, we
can write down the character of the Verma module for $(\vec \ell,\vec
m,\vec s)$ as
\begin{eqnarray*}
 f_{\a}(\tau)
=\frac{\Th_{s_0,2}(\tau)}{\eta(\tau)}\frac{\Th_{m_0,KJ}(\tau)}{\eta(\tau)}
\chi_{m_1}^{\ell_1,s_1}(\tau)\dots \chi_{m_r}^{\ell_r,s_r}(\tau).
\end{eqnarray*}

As a consistency condition, we must impose a modular invariance on the 
total partition function, and moreover we may use only the
states with integral U(1) charges since we want a Calabi-Yau CFT.
For this purpose, we will write down the transformation laws of 
the characters $f_\a$ under the modular transformations
 $\tau\rightarrow \tau +1$ and $\tau\rightarrow -1/\tau$
\begin{align*}
 &f_\a(\tau+1)=\e{-\frac12 (\vec m \bullet \vec m+
\vec s \bullet \vec s)}f_\a(\tau),\\
 &f_\a(-1/\tau)=\sum_{\a'}^{\rm even}S_{\a\a'}f_{\a'}(\tau),\\
 &S_{\a\a'}:=A_{\vec \ell \vec \ell'}
\left(\prod_j\frac{1}{\sqrt{8N_j}}\right)\frac{1}{\sqrt{8KJ}}
\e{\vec m \bullet \vec m'+\vec s \bullet \vec s'},\\
 & A_{\vec \ell \vec \ell'}:=\prod_{j=1}^{r} A_{\ell_j\ell_j'}
=\prod_{j=1}^{r}\sqrt{\frac{2}{N_j}}
\sin \pi \frac{(\ell_j+1)(\ell_j'+1)}{N_j},\\
 & \vec m \bullet \vec m':=
-\frac{m_0m_0'}{2KJ}+\sum_{j=1}^r\frac{m_jm_j'}{2N_j},\\
 & \vec s \bullet \vec s':=-\frac{s_0s_0'}{4}-\sum_{j=1}^{r}
\frac{s_js_j'}{4}.
\end{align*}
To make a modular invariant partition function with these states, 
we further introduce a special vector
$\vb$ in the same type of $\vec m$
\begin{align*}
 \vb=(-2J,2,\dots,2).
\end{align*}
Then, the charge integrality condition can be expressed as the following
``beta constraint''
\begin{align}
  -\frac{m_0}{K}+\sum_{j=1}^{r}\frac{m_j}{N_j}=\vb\bullet\vec m
\in \Zb\,. \label{BetaCondition}
\end{align}
With these notations, we write down 
the NS-sector partition function as a combination of 
characters
\begin{align}
\Tr_{\rm NSNS}\left[q^{L_0-\frac{\ch}{8}}\bar q^{\bar L_0-\frac{\ch}{8}}\right]
=\frac{1}{\sqrt{\tau_2}|\eta(\tau)|^2}
\sum_{b\in \Zb_{K}}
\sum_{\substack{\vec \ell,\vec m,\\ \vec s,\vec{\bar s}=0,2,\\
\ell_j+m_j\equiv 0 \mod 2, \\
\vb\bullet \vec m \in \Zb}}
f_{(\vec \ell,\vec m,\vec s)}(\tau)
\bar f_{(\vec \ell,\vec m+b \vb,\vec{\bar s})}(\taub).
\label{ClosedStringNSAmplitude}
\end{align}
Similarly the RR-sector counterpart 
can also be calculated as
\begin{eqnarray}
&&\Tr_{\rm RR}\left[(-1)^{F}
q^{L_0-\frac{\ch}{8}}\bar q^{\bar L_0-\frac{\ch}{8}}\right]\non
&&\qquad =\frac{1}{\sqrt{\tau_2}|\eta(\tau)|^2}\sum_{b\in \Zb_{K}}
\sum_{\substack{\vec \ell,\vec m,\\ \vec s,\vec{\bar s}=1,3,\\
\ell_j+m_j\equiv 1 \mod 2,\\
\vb\bullet \vec m \in \Zb}}
f_{(\vec \ell,\vec m,\vec s)}(\tau)
\bar f_{(\vec \ell,\vec m+b \vb,\vec{\bar
 s})}(\taub)(-1)^{\sum_j(s_j-\bar s_j)/2+b}.\nn\\
\label{ClosedStringWittenIndex}
\end{eqnarray}
It is nothing but a Witten index of the model and 
geometrical properties of the target manifolds are 
encoded in this formula.

We can check that these partition functions actually satisfy the right
modular properties and construct consistent string theories propagating
in singular target manifolds.  The picture in this section is based on
the closed strings on the singular manifolds. However D-branes couple
with strings only through open strings and it is important to develop a
method to describe open strings in these models.  The open strings can
end on susy cycles of the target manifolds and encode information on
homology cycles of the manifolds.  In the context of CFT, we are able to
analyze properties of these boundaries of the open strings based on the
boundary states.  In the next section, we will construct boundary states
associated with these singular manifolds and investigate their
properties.

\section{Boundary states and the intersection form}
\subsection{The open string in the $\Ncal=2$ Liouville theory}
Here we consider  D-branes in our model and
analyze the properties of the associated open strings. 
Generally it is known that 
two types (A-type and B-type) of boundary conditions are possible 
for $\Ncal =2$ superconformal currents.
These boundary conditions in the open string channel 
are defined on the worldsheet boundary $z=\bar{z}$;
\begin{itemize}
 \item A-type boundary condition
     \begin{align*}
         T=\bar T,\ G^{\pm}=\ve \bar G^{\mp},\ J=-\bar J.
     \end{align*}
 \item B-type boundary condition
     \begin{align*}
         T=\bar T,\ G^{\pm}=\ve \bar G^{\pm},\ J=+\bar J.
     \end{align*}
\end{itemize}
Here, $\ve=1$ in R-sector with $z>0$ and NS-sector, $\ve=-1$ in
R-sector with $z<0$ .
First we consider 
the boundary conditions in the $\Ncal=2$ Liouville sector.
Using $\Ncal=2$ superconformal currents represented by free fields
(\ref{LiouvilleSCA}), 
we express the boundary conditions for free fields
 $\phi$, $Y$ and $\psi^{\pm}$;
\begin{itemize}
 \item A-type
\begin{align*}
 \del\phi=\bar\del\phi\text{ (Neumann) },\ 
       \del Y=-\bar\del Y\text{ (Dirichlet) },\ 
       \psi^{\pm}=\varepsilon\bar\psi^{\mp}.
\end{align*} 
\item B-type
\begin{align*}
       \del\phi=\bar\del\phi\text{ (Neumann) },\ 
       \del Y=+\bar\del Y\text{ (Neumann) },\ 
       \psi^{\pm}=\varepsilon\bar\psi^{\pm}.
\end{align*}
\end{itemize}
We make a remark here;
We don't consider a Dirichlet boundary condition
for the linear dilaton field $\phi$ because it has a subtle problem.
For an example, if we naively take a boundary condition of $\phi$ as
$\del \phi=-\bar\del \phi$, then, from the form of the current 
(\ref{LiouvilleSCA}), we can show that the boundary condition on the 
stress tensor does not satisfy $T=\bar T$. So, we conclude that
$\del \phi=-\bar\del \phi$ is not a good boundary condition.

In the previous section, we consider closed strings on the Calabi-Yau 
manifold. In that case, modular transformations play essential roles in 
constructing consistent string theories.
In the open string case, important information on modular 
transformations are encoded in the annulus amplitudes.
So we will consider the annulus amplitudes 
and study modular properties of them. 

It is well-known that
the annulus amplitude is calculated either as an open string
1-loop partition function or as a closed string transition amplitude
from one boundary state to the other. We denote the moduli parameter
of the annulus by $\tau_2$. It is the radius of the circular direction 
of the annulus when we
normalize 
the length of the perimeter to $\pi$. We also use the following notations in
the open string channel
\begin{align*}
 \tau=i\tau_2,\ q=e^{2\pi i \tau}.
\end{align*}
On the other hand, 
we use the following notations in the 
closed string channel
\begin{align*}
\taut=i\taut_2=-1/\tau,\ \qt=e^{2\pi i \tilde{\tau}}\,,
\end{align*}
because it is more useful to set
the radius of the circle to be $1$
in the closed string case. Then
the length of the segment
becomes $\pi \tilde \tau_2=\pi/\tau_2$. 

Now we introduce a boundary state $|B_L\rra$ for the 
linear dilaton. It is determined uniquely 
because the boundary condition on the linear dilaton is always
Neumann type and has no free parameters.
We can easily calculate the annulus amplitude by open string channel
\begin{align}
\Tr_{O}q^{L_0^O-(1+3Q^2)/24}=\frac1{\sqrt{\tau_2}\eta(\tau)}=
\lla B_L|\tilde q^{L_0^{C}+\bar L_0^{C}-(1+3Q^2)/12} |B_L\rra\,.
\end{align}
Because the linear dilaton sector has no contribution to the
U(1) charge, and because we treat it as a free field,
it decouples from the other sector
\footnote{Very recently, a paper \cite{ES0011} appeared where 
a boundary state and related amplitudes in the Liouville sector are 
discussed based on the perturbative expansions of screened vertex
operators and their analytic continuations.
}, as follows. If we denote the boundary states of the other sectors
than linear dilaton, i.e. the free fermions, the $S^1$ boson, and the
minimal models, $|\alpha\rra$, we can write the total boundary state
as $|B_L\rra\otimes|\alpha\rra$. The annulus amplitude can be written as
\begin{align*}
& (\lla B_L|\otimes\lla \tilde\alpha|)
\tilde q^{L^{\text{(total)}}_0{}^{C}+\bar L^{\text{(total)}}_0{}^{C}
-c/12} (|B_L\rra \otimes|\alpha\rra)\\
&\qquad=\lla B_L|\tilde q^{L^{\text{(linear dilaton)}}_0{}^{C}+
\bar L^{\text{(linear dilaton)}}_0{}^{C}-(1+3Q^2)/12} |B_L\rra\
\lla\alpha|\tilde q^{L^{\text{(other)}}_0{}^{C}+\bar L^{\text{(other)}}_0{}^{C}
-c_{\text{other}}/12}|\alpha\rra\\
&\qquad=\frac1{\sqrt{\tau_2}\eta(\tau)}\lla\alpha|\tilde q^{L^{\text{(other)}}_0{}^{C}+\bar L^{\text{(other)}}_0{}^{C}-c_{\text{other}}/12}|\alpha\rra.
\end{align*}
In the rest of this paper, we neglect the linear dilaton factor, but
we can always obtain the full amplitude by multiplying the linear dilaton
factor $\frac1{\sqrt{\tau_2}\eta(\tau)}$ .

As a second case, we look at the $S^1$ sector.
The boundary states in the bosonic $S^1$ sector are expressed by
ordinary coherent states. For the A-type (Dirichlet) boundary condition, 
the associated state is expressed as
\begin{align*}
  |p\rra_{A}:= \exp\left[\sum_{n=1}^{\infty}\frac1n Y_n \bar Y_n\right]
|p,\bar p=p\ra,
\end{align*}
where $|p,\bar p\ra$ is the closed string Fock vacuum with zero mode
eigenvalues $p$ and $\bar p$.
For the B-type (Neumann) boundary condition, we can obtain the boundary
state by changing signs of $\bar{Y}_n$, $\bar{p}$ for the A-type case
\begin{align*}
  |p\rra_{B}:= \exp\left[-\sum_{n=1}^{\infty}\frac1n Y_n \bar Y_n\right]
|p,\bar p=-p\ra.
\end{align*}
In the following discussions, 
we sometimes omit the subscripts $A,B$ 
when the methods of calculations are applicable
in both types of states.

With these boundary states, 
the transition amplitude is evaluated for the $S^1$ sector
by using the Dedekind eta function
\begin{align*}
  \lla p' | \tilde q^{H^C}| p \rra=\delta_{p,p'}
\frac{\tilde q^{\frac12p^2}}{\eta(\taut)}.
\end{align*}
But this amplitude does not have good properties under  
modular transformations. In order to improve this defect, 
we take a linear combination of 
$| p \rra$'s in the same manner
as that in the closed string ``Verma module'' case
and define a state $|m_0\rra $ 
\begin{align*}
  |m_0\rra =\sum_{u\in \Zb}|p=\frac{2KJu+m_0}{KQ}\rra.
\end{align*}
In this case, an associated transition amplitude between 
$|m_0\rra $ and $|m_0\rra' $ are 
expressed as a combination of a theta function and the eta function
\begin{align*}
  \lla m_0'| \tilde q^{H^C} | m_0 \rra :=\deltam{m_0-m_0'}{2KJ}
\frac{\Th_{m_0,KJ}(\tilde\tau)}{\eta(\tilde\tau)}.
\end{align*}
It has nice properties under modular transformations and 
we take this as a candidate of constituent block of a total 
boundary state.
In the next subsection, we collect results about boundary states 
of various constituent fields
discussed in this section and analyze a total boundary state of 
the whole theory.

\subsection{Total boundary states and the intersection form}
In this subsection we consider the whole theory realized as a product of 
various models
\begin{align*}
  \Rb_{\phi}\times S^1\times M_{N_1}\times M_{N_2}\times \dots \times M_{N_r}
\end{align*}
and construct associated boundary states.

In order to complete this program, we still have to make boundary states 
associated with minimal models. It seems difficult to 
construct full boundary states associated with the tensor product of 
minimal models. In this paper, we shall concentrate on 
``rational boundary states''. 
They are constructed as (linear combinations of) tensor products of 
boundary states of sub-theories 
and are expressed as (omitting the $\Rb_{\phi}$ part)
\begin{align*}
  |\a\rra
:=|s_0\rra\otimes |m_0\rra \otimes |\ell_1,m_1,s_1\rra\otimes
\dots \otimes |\ell_r,m_r,s_r\rra.
\end{align*}
Here the index $\a$ is the same symbol as that used in the section
\ref{ClosedString}, and $|\ell_j,m_j,s_j\rra$'s are ordinary Ishibashi
states of minimal models\cite{RS9712}.

The choices of boundary types lead to make differences in 
allowed states.
For the A-type boundary state, the U(1)
charges of the left and right movers are the same 
and all the A-type boundary
states with the condition (\ref{BetaCondition}) are available.
On the other hand, for the B-type boundary condition, the U(1)
charges of the left and right movers have the same absolute values but
opposite signs and the allowed
states must satisfy a condition
\begin{align*}
 \vec m=\frac12 b\vb,\qquad b\in\Zb.
\end{align*}
Using above Ishibashi states, we can construct the Cardy states 
with appropriate coefficients determined by 
the S matrix under the $S$ modular transformation. 
In the next two subsections, we construct the Cardy states
concretely for A-type and B-type cases. 

\subsubsection{A-type boundary states}
A Cardy state\cite{Car89} with the A-type boundary condition is labelled by
a set of indices $\alpha=(\Vec L,\Vec M, \Vec S)$. Here 
the $\Vec L,\Vec M, \Vec S$ are respectively the same type vectors as 
$\vec \ell,\vec m, \vec s$. The Cardy state
 $|\alpha\rra$ is defined as a linear combination of Ishibashi states
\begin{align}
  & |\alpha\rra_A =\frac{1}{\kappa^A_{\alpha}}
 \sum_\a^{\rm beta}B_{\alpha}^{\a}|\a\rra_A ,\nn\\
 & B_{\alpha}^{\a}=\frac{S_{\alpha \a}}{\sqrt{S_{\0\a}}},\nn
\end{align}
where the symbol $\0:=(\vec\ell=0,\vec m=0,\vec s=0)$ is introduced.
The normalization constant $\kappa^A_{\alpha}$ is determined
so that the states $|\alpha\rra_A$ satisfy Cardy conditions
of the cylinder amplitudes.

First we calculate an NS-sector amplitude between two of the Cardy states,
$|\alpha\rra_A$ and ${_A}\lla \tilde \alpha|$. From the viewpoint 
of open string channel, 
this amplitude should be equal to an NS-sector 
1-loop partition function of the open string
\begin{align*}
 Z^{A}_{\alpha\tilde\alpha}
&={}_A\lla \tilde \alpha | \tilde q^{H^C}|\alpha\rra_A{}_{\rm\  NS} \\
&=\frac{1}{\kappa^A_{\alpha}\kappa^A_{\tilde\alpha}}
\sum_{\a,\tilde \a}^{\rm beta,NS}
B_{\alpha}^{\a}B_{\tilde\alpha}^{\tilde \a}{}^* 
\lla \tilde \a | \tilde q^{H^C} | \a \rra \\
&=\frac{1}{\kappa^A_{\alpha}\kappa^A_{\tilde\alpha}}
\sum_a^{\rm beta,NS}B_{\alpha}^{\a}B_{\tilde\alpha}^{\a}{}^*
f_\a(\tilde q)\\
&=\frac{1}{\kappa^A_{\alpha}\kappa^A_{\tilde\alpha}}\sum_{\a'}^{\rm even}
\sum_\a^{\rm beta,NS}B_{\alpha}^{\a}B_{\tilde\alpha}^{\a}{}^*
S_{\a\a'}
f_{\a'}(q).
\end{align*}
The symbol $\dis \sum^{\rm beta}$ means the sum is taken under the beta
constraint (\ref{BetaCondition}) we discussed in the previous section.
Also the terms $B_{\alpha}^{\a}B_{\tilde\alpha}^{\a}{}^* S_{\a\a'}$ are
reexpressed as
\begin{align*}
 B_{\alpha}^{\a}B_{\tilde\alpha}^{\a}{}^* S_{\a\a'}
&=\frac{S_{\alpha \a}S_{\tilde \alpha \a}^* S_{\a\a'}}{S_{\0\a}}\\
&=\frac{A_{\vec L \vec \ell}A_{\vec{\tilde L} \vec \ell}
A_{\vec \ell \vec \ell'}}{A_{\vec 0 \vec \ell}}
\left(\prod_j\frac{1}{8N_j}\right)\frac{1}{8KJ}
\e{\vec m \bullet(\vec M -\vec{\tilde M}+\vec m')
+\vec s \bullet (\vec S-\vec{\tilde S}+\vec s')}.
\end{align*}
To evaluate the sum $\dis \sum_\a^{\rm beta,NS}$ 
under the beta constraint, we introduce a Lagrange
multiplier $\nu_0$
and rewrite the above sum as
\begin{align*}
 \sum_{\vec \ell,\vec m}^{\rm beta,NS}=\sum_{\vec \ell,\vec m}^{\rm even
 ,NS} \frac{1}{K}\sum_{\nu_0=0}^{K-1}\e{\nu_0\vb\bullet\vec m}.
\end{align*}
Thus we obtain the partition function of the 
open string in the NS-sector
\begin{align*}
 Z^{A}_{\alpha\tilde\alpha}=\frac{1}{\xi_{\alpha}\xi_{\tilde\alpha}}
\sum_{\a'}^{\rm even,NS}
\sum_{\nu_0=0}^{K-1}\left(\prod_j N_{L_j\tilde L_j}^{\ell_j'}\right)
\delta_{\vec M -\vec{\tilde M}+\vec m'+\nu_0 \vb} f_{\a'}(q),
\end{align*}
where $\xi_{\alpha}$ is a constant related with $\kappa^{A}_{\alpha}$.
The symbols $N_{L_j\tilde L_j}^{\ell_j'}$'s represent 
the SU(2) fusion coefficients and their explicit formulas are 
shown in the appendix A. If we want the total partition function
including the linear dilaton sector, we should multiply the
factor $\frac{1}{\sqrt{\tau_2}\eta(\tau)}$.

Now, let us calculate a kind of topological invariants, 
the ``open string Witten index''. 
This index has information on the intersection pairings
between two sheaves in the geometric language and 
we compare these results with those obtained by 
purely geometrical techniques.

The open string Witten index can be calculated 
in the closed string channel by evaluating the RR amplitude between
$|\alpha\rra_A$ and ${_A}\lla \tilde \alpha|$ with an insertion of
$(-1)^{F_L}$.
\begin{align*}
  I^A_{\alpha\tilde\alpha}
:=\Tr_{\alpha,\tilde\alpha,\rm R}(-1)^{F}q^{H^O}=
{}_A\lla \tilde \alpha |(-1)^{F_L} \tilde q^{H^C}|\tilde \alpha 
\rra_A{}_{\rm\  R}.
\end{align*}
Using the Cardy states we obtained, 
an associated open string Witten index is expressed as
\begin{align*}
 I^A_{\alpha\tilde\alpha}&=\frac{1}{\kappa^A_{\alpha}\kappa^A_{\tilde\alpha}}
\sum_{\a,\tilde \a}^{\rm beta,R}
B_{\alpha}^{\a}B_{\tilde\alpha}^{\tilde \a}{}^* 
\lla \tilde \a |  (-1)^{F_L}\tilde q^{H^C}| \a \rra \\
&=\frac{1}{\kappa^A_{\alpha}\kappa^A_{\tilde\alpha}}
\sum_\a^{\rm beta,R}B_{\alpha}^{\a}B_{\tilde\alpha}^{\a}{}^*
(-1)^{-\frac12 s_0-\sum_j\frac12 s_j+\vb\bullet\vec m}
f_\a(\tilde q)\\
&=\frac{1}{\xi_{\alpha}\xi_{\tilde\alpha}}
\sum_{\a'}^{\rm even,R}\sum_{\nu_0=0}^{K-1}
\left(\prod_{j=1}^{r} N_{L_j\tilde L_j}^{\ell_j'}\right)
\delta_{\vec M -\vec{\tilde M}+\vec m+(\nu_0+1/2) \vb}
(-1)^{\sum_{j=0}^{r}\frac{S_j-\tilde S_j+s_j}{2}} f_{\a'}(q).
\end{align*}
By using the fact that only the ground states contribute to the Witten index,
we obtain a concrete formula for this
\begin{align*}
 I^A_{\alpha\tilde\alpha}
&=\frac{1}{\xi_{\alpha}\xi_{\tilde\alpha}}
(-1)^{ \sum_{j=0}^{r}\frac{S_j-\tilde S_j}{2}}
\sum_{\nu_0=0}^{K-1}
\left(\prod_{j=1}^r N_{L_j\tilde L_j}^{2\nu_0+M_j-\tilde M_j}\right)
\deltam{-(2\nu_0+1)J+M_0-\tilde M_0}{2KJ}
\frac{\theta_1(\tau)}{(\eta(\tau))^2}\\
&=\frac{1}{\xi_{\alpha}\xi_{\tilde\alpha}}
(-1)^{ \sum_{j=0}^{r}\frac{S_j-\tilde S_j}{2}}
\left(\prod_{j=1}^r N_{L_j\tilde L_j}^{\frac{M_0-\tilde M_0}{J}
+M_j-\tilde M_j-1}\right)
\deltam{M_0-\tilde M_0-J}{2J}
\frac{\theta_1(\tau)}{(\eta(\tau))^2}.
\end{align*}
Actually
this index vanishes because of
the relation $\theta_1(\tau)=0$.
Even if we multiply the linear dilaton factor
 $\frac{1}{\sqrt{\tau_2}\eta(\tau)}$ , the index still vanishes.
In the next subsection, we construct a B-type boundary states
associated with this singular manifold.

\subsubsection{B-type boundary states}

In contrast to the A-type boundary condition, 
available Ishibashi states
are restricted to those with $\vec m=(1/2) \ b \vb \ \ ,b\in \Zb$ 
for the B-type case. 
Then, a Cardy state of the B-type boundary condition is
defined as a linear combination of Ishibashi states 
\begin{align}
   & |\alpha\rra_B =\frac{1}{\kappa^B_{\alpha}}
 \sum_{\a;\vec m=\frac12 b \vb}^{\rm beta }B_{\alpha}^{\a}|\a\rra ,\nn\\
 & B_{\alpha}^{\a}=\frac{S_{\alpha \a}}{\sqrt{S_{\0\a}}}.\nn
\end{align}
Since $B_{\alpha}^{\a}$ depends on the vector $\Vec M$ only through 
the combination $b\vb \bullet
\Vec M$, the Cardy state $|\alpha\rra_B $ is labelled by
a number $M:=\vb\bullet\Vec M$ and vectors $\Vec L, \Vec S$.
By applying the similar methods in the A-type case, 
the NS-sector partition function, equivalently, the NS closed string amplitude
from $|\alpha\rra_B$ to ${}_B\lla \tilde \alpha|$ are evaluated as
\begin{align}
 Z_{\alpha\tilde\alpha}^{B}&={}_B\lla \tilde \alpha | \tilde q^{H^C}
|\alpha \rra_{B \ \rm NS} \nn\\
 &= \frac{1}{\zeta_{\alpha}\zeta_{\tilde\alpha}}
\sum_{\a'}^{\rm even,NS}\deltam{\frac12(M-\tilde M +
 K\vb\bullet \vec m')}{K} \prod_{j=1}^rN^{\ell_j'}_{L_j,\tilde L_j}
f_{\a'}(q),
\label{B-NSAmplitude}
\end{align}
where $\zeta_{\alpha}$ is a constant proportional to
$\kappa^B_{\alpha}$.
Also this formula leads us to calculate the open string Witten index 
\begin{align}
  I^B_{\alpha\tilde\alpha}&=
{}_B\lla \tilde \alpha |  (-1)^{F_L}\tilde q^{H^C}| \alpha \rra_B{}_{\rm
\  R} \nn \\
&=\frac{1}{\zeta_{\alpha}\zeta_{\tilde\alpha}}
(-1)^{\frac{S-\tilde S}2} \sum_{m_1',\dots, m_R'}
\deltam{\frac12\left[M-\tilde M+\sum_j\frac{K(m_j'+1)}{2N_j}\right]}{K}
\left(\prod_{j=1}^r N_{L_j,\tilde L_j}^{m_j-1}\right)\frac{\theta_1(\tau)}{(\eta(\tau))^2}.
\label{B-Index}
\end{align}
This index vanishes, even if we include the linear dilaton factor,
for the same reason as that in the A-type case.

In this and the previous subsections, we construct concretely boundary
states for the A-,B-type cases.
In the next subsection, we take a simple class of manifolds 
as an example and 
calculate its open string Witten index.
That is compared with a 
geometric result in the next section and confirms the 
validity of our results about boundary states.

\subsection{A simple class of manifolds}
In the previous subsection, we showed that the open string Witten index 
vanishes. But it is factorized
into $\frac{\theta_1(\tau)}{\sqrt{\tau_2}\eta(\tau)^3}\ (=0)$ and a 
rather nontrivial factor. In this subsection, we take a simple
class of manifolds for an example and consider the meaning of 
the nontrivial factor.

\begin{figure}
 \begin{center}
  \includegraphics[width=5cm]{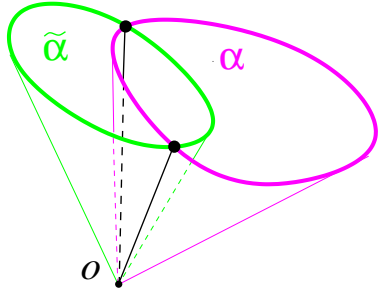}
 \end{center}
 \caption{The image of the intersecting noncompact two cycles. The whole
 space expresses the singular Calabi-Yau manifold $X$ and the origin $O$
 is the singularity. The radial direction expresses the $\phi$ and $Y$
 direction, along which all the B-type D-branes spread. The cones
 $\alpha$ and $\tilde \alpha$ express the D-branes. Since they intersect
 at a surface $\Cx$ of $\phi$ and $Y$, the intersection number is
 indefinite. This is the reason why we got the intersection number
 $0$. If we restrict to the section of the cycles (the base of the
 cone), the intersection number of this section in the manifold $X/\Cx$
 is possibly definite. We claim that the nontrivial factor of the open
 string Witten index corresponds to this intersection pairing in $X/\Cx$
 } \label{Cycles}
\end{figure}

For simplicity, 
we concentrate on the B-type boundary states. The singular manifold $X$
is realized as a fibered space over $X/\Cx $\cite{Yam0007}
\begin{align*}
 X=X/\Cx\times_{\mathrm{f}}\Cx,
\end{align*}
where the symbol ``$\times_{\mathrm{f}}$'' means a fibration. 
In the Gepner-like description,
$X/\Cx$ seems to correspond to a direct product of the 
minimal models and $\Cx$ seems to be related with
 the $\Ncal=2$ Liouville part. For the B-type D-branes,
the boundary conditions on two bosons $\phi$, $Y$ in the 
$\Ncal=2$ Liouville part 
are Neumann types. In other words, all the B-type D-branes spread to 
the $\Cx$ direction. The $\Cx$ part has a trivial structure and
the B-type D-brane is essentially characterized by cycles of $X/\Cx$
as we show in Fig. \ref{Cycles}.

Let us consider a simple case that $X$ is written as
\begin{align*}
 z_1^N+z_2^N+\dots+z_{n+1}^N=0 \text{ in } \Cb^{n+1}.
\end{align*}
When one notes the fact $\Cb P^n\cong (\Cb^{n+1}-\{0\})/\Cx$, 
the $X/\Cx$ in our case turns out to be a Ricci positive $(n-1)$-fold in 
$\Cb P^n$.
Moreover, we concentrate to the boundary states with $L=S=0$ to
simplify our analyses.
Also we may set $\zeta_{\alpha}=1$ in Eq.(\ref{B-NSAmplitude}) 
because the Cardy condition is satisfied even for that case.
In this case, the boundary states are labelled only
by the index $M$. 
Then we can write down a 
nontrivial factor of the open string Witten index in the B-type case
explicitly
\begin{align*}
 \hat I_{M\tilde M}&=\sum_{m_1',\dots, m_r'=0,\dots,2N}
\deltam{M-\tilde M+\sum_j(m_j'+1)}{2N}
\left(\prod_{j=1}^{r} N_{0,0}^{m_j-1}\right)\\
&=\sum_{m_1',\dots, m_r'=0,\dots,2N}
\deltam{M-\tilde M+\sum_j(m_j'+1)}{2N}
\prod_{j=1}^{r}\left(\delta_{m_j',1}-\delta_{m_j',2N-1}\right)\,.
\end{align*}
Because $M$ and $\tilde M$ should be even integers, we can introduce
integers $a,b$ ($a,b=0,1,\dots,N-1$) as
$a=M/2$ and $b=\tilde M/2$.
The $\hat I$ is represented in a compact formula
\begin{align*}
 \hat I_{a,b}&=\sum_{c=0}^{r}\deltam{a-b+c}{N}\binom{r}{c}(-1)^{r-c}\\
&=(-1)^r\sum_{m\geq 0}\binom{r}{b-a+mN}(-1)^{b-a+mN}\,.
\end{align*}
It depends on only variables $a$ and $b$ and is interpreted as 
a $N\times N$ matrix $I_G$
\ba 
&&I_G=(g^{-1}-1)^r\,,\,\,\,g^N=1\,.\label{gmodel}
\ea
Here the shift matrix $g$ is represented as an 
matrix with $N \times N$ entries
\ba
g=\left(
\gyoretu{
0&1& & & & \cr
 &0&1& & & \cr
 & &0&1& & \cr
 & & &\ddots &\ddots  & \cr
 & & & &0&1 \cr
1&0&0&\cdots &0 &0\cr}
\right)
\,.\nom
\ea

The pairing obtained here is neither symmetric
nor anti-symmetric with respect to two indices. 
We discuss the (anti-)symmetric part of
this pairing.
We consider the (anti-)symmetric part of $I_G$ as
\begin{align*}
 \IL:=-[I_G^{t}+(-1)^{r-1}I_G],
\end{align*}
then, $\IL$ can be written as
\begin{align}
  \IL=(g^{-1}-1)^r(g^{r-N}-1).\label{SymmetricForm}
\end{align}
This is interpreted as follows.

In an ordinary compact Gepner model written by the Landau-Ginzburg
model with a superpotential
\begin{align*}
 W=z_0^{N_0}+z_1^{N_1}+\dots+z_r^{N_r},
\end{align*}
the intersection form of B-type cycles becomes\cite{BDLR9906,DR9910}
\begin{align*}
 I_{L=\tilde L=0}=\prod_j(g^{-K/N_j}-1),\quad K=\lcm(N_j).
\end{align*}

According to the paper \cite{Ler0006}, if 
we formally apply this formula to our case with a negative and
fractional power part
\begin{align*}
 W=z_0^{-N/(r-N)}+z_1^N+\dots+z_r^N,
\end{align*}
then (\ref{SymmetricForm}) is obtained.

We compare these results with a 
calculation of an intersection pairing in terms of geometrical methods 
in the next section.

\newcommand{\X}{V}
\section{Geometric interpretation}
In this section, we analyze topological properties of 
the manifold $M=X/\Cx$ discussed in the previous section. 
The study here is based on geometrical methods and 
we compare results of intersection pairings obtained from 
two different approaches. It confirms the validity of 
our boundary states we constructed in the CFT.

\subsection{Intersection pairing}
In this section, we interpret the results in the Gepner model 
geometrically. 
First we take a Ricci positive $d=n-1$ dimensional manifold $M=X/\Cx$
\ba
&&M\,;\,z_1^N+z_2^N+\cdots +z_r^N=0\qquad \mbox{in}\,\,\,\Cb P^{r-1}\,,\non
&&r=d+2=n+1\,.\nom
\ea
Its first Chern class is evaluated by using a 
cohomology element $H\in H^2(M)$ 
\ba
c_1(TM)=(r-N)H\,.\nom
\ea
Because M is realized as a zero locus of an ambient space $\X=\Cb P^{r-1}$, 
we can discuss topological properties of $M$ by analyzing 
characteristic classes of $\X$ through a restriction on $M$. 
Our results in the previous sections are based on the 
analyses of open strings in the Gepner model and 
the associated objects ``D-branes''(susy cycles) are expected to 
play important roles in our theory.
A suitable basis for D-branes in the orbifold 
point\cite{DD0006,Sug0003A,Hos0007,GJ0010,Tom0010,May0010} is a 
set of line bundles (coherent sheaves) 
$\{{\cal O}(-a)\}$ $(a=0,1,2,\cdots ,r-1)$ over $\X$.
(Associated analyses based on the Landau-Ginzburg models are 
performed  in papers\cite{HIV0005,GJS9907,GJ0003,GJS0007}. Also 
applications to ${\cal N}=2$ gauge theories are proposed 
in papers\cite{LLS0006,Ler0006,FKLLSW0007}.)
The cylinder amplitude is interpreted as an index of the Dirac operator 
with boundary gauge bundles $(E,E')$. In other words, it is 
a natural inner product on these bundles and is expressed as a 
relative Euler characteristic $\chi_{\X}(E,E')$ over $\X$ (for $E=\co
(-a)$, $E'=\co (-b)$)
\ba
&&\langle \co (-a),\co (-b)\rangle_\X:=\chi_{\X}(\co (-a),\co (-b))\non
&&\qquad =\int_{\X}\cha (\co (-a)^{\ast})\cha (\co (-b))\td (T\X)\non
&&\qquad =
\left(
\gyoretu{r-1+a-b\\ a-b}
\right)=:I_{a,b}
\,,\label{chi}\\
&&\qquad\qquad (0\leq a \leq r-1\,;\,0\leq b\leq r-1)\,,\non
&&\td (T\X)=\left(\frac{H}{1-e^{-H}}\right)^r\,.\nom
\ea
{}From now on, we use an abbreviated notation 
$R_a$ for ${\cal O}(-a)$ 
$(a=0,1,\cdots ,r-1)$.

Next we construct the dual basis $\{S^{a}\}$ of the $\{R_{a}\}$ 
$(0\leq a\leq r-1)$ as
\ba
&&\cha ({S^a})^{\ast}:=\sum_b (I^{-1})_{a,b} \cha (R_b)^{\ast}
=\sum_b (-1)^{a-b}\left(\gyoretu{r\cr a-b}\right)\cha (R_b)^{\ast}
\,,\nom
\ea
They satisfy orthonormal conditions with respect to the 
intersection pairing
\ba
&&\langle S^{a},R_{b}\rangle_\X =
\chi_\X (S^a,R_b)=\chi_\X ({S^a}^{\ast}\otimes R_b)=
\delta^{a}{}_{b}\,.\nom
\ea
The set of line bundles $\{{\cal O}(a)\}$ $(a=0,1,\cdots ,r-1)$
is a strongly exceptional collection of the $\X=\Cb P^{r-1}$ 
and turns out to be a foundation of an associated helix of $\X$.
We can introduce an operation ``left mutation ${\bf L}$'' on the 
set $\{{\cal O}(a)\}$ as exact sequences
\ba
0\rightarrow \mbox{Ext}^0(\co (a-1),\co (a))\otimes \co (a-1)
\rightarrow \co (a)\rightarrow {\bf L}_{\co (a-1)}(\co (a))
\rightarrow 0\,,\label{exact1}
\ea
or
\ba
0\rightarrow {\bf L}_{\co (a-1)}(\co (a))
\rightarrow \mbox{Ext}^0(\co (a-1),\co (a))\otimes \co (a-1)
\rightarrow \co (a)
\rightarrow 0\,.\label{exact2}
\ea
Here we used a condition $\mbox{Ext}^0(\co (a-1),\co (a))=
H^0(\co (a-1),\co (a))\neq 0$ for $\X$.
It induces a relation of Chern characters of the bundles 
\ba
\pm \cha ({\bf L}_{a-1}\co (a))=
\cha (\co (a))-\chi_\X(\co (a-1),\co (a))\cha (\co (a-1))\,,\label{mutation}
\ea
where we introduced a notation 
${\bf L}_{a-1}\co (a)={\bf L}_{\co (a-1)}\co (a)$. 
The sign in Eq.(\ref{mutation}) depends on the choice of sequences 
Eqs.(\ref{exact1}),(\ref{exact2}), that is to say, 
$(+)$ for Eq.(\ref{exact1}), 
$(-)$ for Eq.(\ref{exact2}).
By using Eq.(\ref{mutation}) iteratively, we can reexpress the 
${S^a}^{\ast}$ as
\ba
\cha (S^a)^{\ast}=\sum^a_{b=0}(-1)^{a-b}\left(\gyoretu{r\cr a-b}\right)
\cha (\co (b))=\cha ({\bf L}_0{\bf L}_1\cdots {\bf L}_{a-1}\co (a))\,.
\ea
Namely, each element of ${S^a}^{\ast}$ can be constructed by acting
left mutations on the $R_a^{\ast}=\co (a)$.

Next we will take an equivalence class $\co (-[a])$ for $\{\co (-a)\}$
because $M$ has an extra cyclic property $\Zb_N$ at the orbifold
point
\ba
\co (-[a+N])=\co (-[a])\,.\nom
\ea
The $[a]$ is defined modulo $N$. Equivalently we can interpret the 
number $a$ as an element of a cyclic group $\Zb_N$.
We shall write these elements of equivalence classes in an abbreviated form 
$R_{[a]}={\cal O}(-[a])$ $(a=0,1,\cdots ,N-1)$.

In the previous sections, we investigate the
 boundary states in the Gepner model.
Generally open strings can end on susy cycles described by 
homology cycles.
When one discusses boundary states, homology classes play essential
roles. 
An important class of topological invariants is an intersection 
pairing of homology cycles. These are related with 
cylinder amplitudes of open strings with boundary gauge bundles or 
sheaves.
Now we define a pairing $\langle R_{[a]} ,R_{[b]}\rangle$ 
on these bundles on $M$
\ba
&&{\bf I}_{a,b}:=
\langle R_{[a]},R_{[b]}\rangle :=\sum_{\ell\geq 0}\sum_{m\geq 0}
\int_M\cha(R_{a+N\ell})^{\ast}\cha(R_{b+Nm})\td (TM)\,,\label{todd}\\
&&\td (TM)=\left(\frac{H}{1-e^{-H}}\right)^r
\left(\frac{1-e^{-NH}}{NH}\right)\,.\nom
\ea
After performing the sum with respect to $m$ in Eq.(\ref{todd}), 
we can rewrite the pairing as
\ba
&&\langle R_{[a]},R_{[b]}\rangle =
\sum_{\ell\geq 0}\int_M\cha(R_{a+N\ell})^{\ast}\cha(R_{b})
\frac{1}{1-e^{-NH}}\cdot
\td (TM)\non
&&\qquad =\sum_{\ell\geq 0}\int_\X\cha(R_{a+N\ell})^{\ast}\cha(R_{b})
\td (T\X)\,.\nom
\ea
Here we used a relation $\int_M(\cdots ) =\int_\X NH\times (\cdots )$.
Thus we obtain an expression for the pairing
\ba
&&\langle R_{[a]},R_{[b]}\rangle = 
\sum_{\ell\geq 0}\int_\X\cha(R_{a+N\ell})^{\ast}\cha(R_{b})
\td (T\X)\non
&&\qquad =:\sum_{c=0}^{r-1}\deltam{a,c}{N}
\int_{\X}\cha (\co (-a)^{\ast})\cha (\co (-b))\td (T\X)
=\sum_{c=0}^{r-1}\deltam{a,c}{N} I_{c,b}\,,\non
&&\qquad\qquad (0\leq a \leq N-1\,;\,0\leq b\leq N-1)\,.\nom
\ea
This definition is natural because 
a usual pairing $I_{a,b}=\langle {\cal O}(-a),{\cal O}(-b)\rangle_\X$
on the set $\{{\cal O}(-a)\}$ $(a=0,1,\cdots ,r-1)$ 
is defined as Eq.(\ref{chi}).
Also the ${\bf I}_{a,b}$ depends on only difference $(a-b)$ and we will write 
this as ${\bf I}_{a-b}$.
The ${\bf I}_{a-b}$\footnote{
Instead of the ${\bf I}_{a,b}$ case, we can construct 
(anti-)symmetric pairings $I^D_{a,b}$ or $I^D_{[a],[b]}$ 
by using an A-roof genus 
\ba
&&I^D_{a,b}:=\int_M\cha (R_a)^{\ast}\cha (R_b)\hat{A}(TM)\,,\non
&&I^D_{[a],[b]}:=\sum_{\ell\geq 0}\sum_{m\geq 0}
\int_M\cha (R_{a+N\ell})^{\ast}\cha (R_{b+Nm})\hat{A}(TM)\,,\non
&&\hat{A}(TM)=e^{-\frac{1}{2}c_1(TM)}\td (TM)\,,\non
&&I^D_{a,b}=(-1)^{r-2}I^D_{b,a}\,,\,\,\,
I^D_{[a],[b]}=(-1)^{r-2}I^D_{[b],[a]}\,.\nom
\ea
However we do not discuss them here.} is  generally neither 
symmetric nor anti-symmetric under exchanges of $a$ and $b$
\ba
{\bf I}_{a-b}=(-1)^r{\bf I}_{b-a+N-r}\,.\nom
\ea
Only for $c_1(TM)=0$ case, the ${\bf I}_{a-b}$ is either symmetric for 
$r=$even or anti-symmetric for $r=$odd.
However we can construct an (anti-)symmetric pairing ${\cal I}_{a,b}$ 
from the ${\bf I}_{a,b}$ 
\ba
&&{\cal I}_{a,b}:={\bf I}_{a,b}+(-1)^{r-1}{\bf I}_{b,a}\non
&&\qquad ={\bf I}_{a-b}+(-1)^{r-1}{\bf I}_{b-a}\,,\label{symmetric}\\
&&{\cal I}_{a,b}=(-1)^{r-1}{\cal I}_{b,a}\,.\nom
\ea
In other words, this ${\cal I}_{a,b}$ can be considered as 
 an (anti)symmetrized version of the ${\bf I}_{a,b}$, that is, 
symmetric for $r=$odd case, anti-symmetric for $r=$even case.
Also the ${\cal I}_{a,b}$ is expressed by using 
characteristic classes
\ba
&&{\cal I}_{a,b}=\sum_{\ell\geq 0}\sum_{m\geq 0}
\int_M\cha(R_{a+N\ell})^{\ast}\cha(R_{b+Nm})\td (TM)\cdot 
(1-e^{-c_1(TM)})\,,\non
&&\qquad =\sum_{\ell\geq 0}\int_\X\cha(R_{a+N\ell})^{\ast}\cha(R_{b})
\td (T\X)\cdot (1-e^{-(r-N)H})\,.\nom
\ea
Next we introduce a dual basis $\{S^{[a]}\}$ of the $\{R_{[a]}\}$ 
$(a=0,1,\cdots ,N-1)$ 
\ba
&&\cha (S^{[a]})^{\ast}=\sum_{c=0}^{r-1}\deltam{c,a}{N}
\cha (S^c)^{\ast}
:=\sum_{\ell \geq 0}
\cha (S^{a+N\ell})^{\ast}\,.\label{s-sheaf}
\ea
They satisfy a set of orthonormal conditions
\ba
&&\langle S^{[a]},R_{[b]}\rangle 
=\chi_{\X}(S^{[a]},R_{[b]})\non
&&\qquad =\sum_{c=0}^{r-1}\deltam{c,a}{N}
\int_{\X}\cha (S^c)^{\ast}\cha (R_b)\td (T\X)\non
&&\qquad =\sum_{c=0}^{r-1}\deltam{c,a}{N}\delta_{c,b}
=\deltam{b,a}{N}=\delta^{a}{}_{b}\,,\non
&&\langle R_{[a]},S^{[b]}\rangle 
=\chi_{\X}(R_{[a]},S^{[b]})\non
&&\qquad =\sum_{c=0}^{r-1}\deltam{c,a}{N}
\int_{\X}\cha (R_c^{\ast})\cha (S^b)\td (T\X)\non
&&\qquad =\sum_{c=0}^{r-1}\deltam{c,a}{N}\delta_{c,b}
=\deltam{b,a}{N}=\delta_a{}^{b}\,,\non
&&\qquad (0\leq a\leq N-1\,;\,0\leq b\leq N-1)\,.\nom
\ea
Then we can evaluate a pairing for each pair of elements 
$S^{[a]}$, $S^{[b]}$ $(a,b=0,1,\cdots ,N-1)$
\ba
&&\langle S^{[a]},S^{[b]}\rangle =\chi_{\X}(S^{[a]},S^{[b]})\non
&&\qquad =\sum_{c=0}^{r-1}\deltam{c,a}{N}
\int_{\X}\cha (S^c)^{\ast}\cha (S^b)\td (T\X)\non
&&\qquad =\sum_{c=0}^{r-1}\deltam{a,c}{N}(I^{-1})_{c,b}
=\sum^{r}_{c=0}\deltam{a,c}{N}
\left(\gyoretu{r\cr c-b}\right)\cdot (-1)^{c-b}\non
&&\qquad =\sum_{m\geq 0}\left(\gyoretu{r\cr
a-b+Nm}\right)(-1)^{a-b+Nm}\label{geom}\,,\\
&&\qquad\qquad (0\leq a\leq N-1\,;\,0\leq b\leq N-1)\,.\nom
\ea
On the other hand, we calculated another intersection matrix $I_G$ 
in Eq.(\ref{gmodel}) of boundary
states in the Gepner model for this case
\ba 
&&M\,;\,z_1^N+z_2^N+\cdots +z_r^N=0\qquad \mbox{in}\,\,\,{\Cb P}^{r-1}\,,\non
&&I_G=(g^{-1}-1)^r\,,\,\,\,g^N=1\,.\nom
\ea
One can express each component of the $I_G{}^t$ in an $(a,b)$th entry
\ba
&&(I_G)^t_{a,b}=(-1)^{r}\sum_{m\geq 0}\left(\gyoretu{r\cr
a-b+Nm}\right)(-1)^{a-b+Nm}\label{gep}\,,\\
&&\qquad (0\leq a\leq N-1\,;\,0\leq b\leq N-1)\,.\nom
\ea
This result Eq.(\ref{gep}) coincides with that in the geometric 
intersection Eq.(\ref{geom}) up to an irrelevant overall sign.
That is to say, the boundary states we have obtained are 
associated with equivalence classes of the bundles (sheaves)
$\{S^{[a]}\}$ $(a=0,1,\cdots ,N-1)$ in Eq.(\ref{s-sheaf}).
However, these are neither symmetric nor anti-symmetric under the 
exchanges $(a,b)\rightarrow (b,a)$.
So we shall consider symmetrized parts of them by using the 
symmetrized pairing in Eq.(\ref{symmetric})
\ba
&&\langle S^{[a]},S^{[b]}\rangle_{\cal I}=
\langle S^{[a]},S^{[b]}\rangle +(-1)^{r-1}
\langle S^{[b]},S^{[a]}\rangle \non
&&\qquad =\sum^{r-1}_{c=0}\deltam{a,c}{N}
\left(\gyoretu{r\cr c-b}\right)\cdot (-1)^{c-b}
+(-1)^{r-1}
\sum^{r-1}_{c=0}\deltam{b,c}{N}
\left(\gyoretu{r\cr c-a}\right)\cdot (-1)^{c-a}
\non
&&\qquad =\sum_{m\geq 0}\left(\gyoretu{r\cr
a-b+Nm}\right)(-1)^{a-b+Nm}
+(-1)^{r-1}
\sum_{m\geq 0}\left(\gyoretu{r\cr b-a+Nm}\right)(-1)^{b-a+Nm}
\,,\non
&&\qquad\qquad (0\leq a\leq N-1\,;\,0\leq b\leq N-1)
\,.\nom
\ea
with $\langle S^{[a]},S^{[b]}\rangle_{\cal I}=(-1)^{r-1}
\langle S^{[b]},S^{[a]}\rangle_{\cal I}$.
Then we can compare this result with the matrix ${\IL}$ and $I_G$ in
Eq.(\ref{SymmetricForm})
\begin{align*}
 &I_G=(g^{-1}-1)^r\,,\non
 &\IL=(g^{-1}-1)^r(g^{r-N}-1)
 =(-1)[I_G+(-1)^{r-1}I_G^t]\,,\non
 &(-1)^{r-1}(\IL)^t_{a,b}=
 (-1)^r[(I_G)^t_{a,b}+(-1)^{r-1}(I_G)_{a,b}]\non
 &=
 \sum_{m\geq 0}\left(\gyoretu{r\cr
 a-b+Nm}\right)(-1)^{a-b+Nm}
 +(-1)^{r-1}
 \sum_{m\geq 0}\left(\gyoretu{r\cr
 b-a+Nm}\right)(-1)^{b-a+Nm}
 \,,\nom
\end{align*}
with $(\IL)^t=(-1)^{r-1}\IL$.
As a conclusion, the $(-1)^{r-1}(\IL)^t_{a,b}$ coincides with the 
$\langle S^{[a]},S^{[b]}\rangle_{\cal I}$ on $\X$
\ba
&&(-1)^{r-1}(\IL)^t_{a,b}=\langle S^{[a]},S^{[b]}\rangle_{\cal I}\,,\non
&&\langle S^{[b]},S^{[a]}\rangle_{\cal I}=(-1)^{r-1}
\langle S^{[a]},S^{[b]}\rangle_{\cal I}\,.\nom
\ea
That is to say, we are able to interpret the symmetrized part of the
pairings of the boundary states as those of the bundles $\{S^{[a]}\}$.
In other words, a state $|\{L=0\};M=2a;S=0\rangle$ corresponds to a 
bundle (sheaf) ${S^{[a]}}^{\ast}$.
Formally an arbitrary state in the Gepner model could be represented as 
some bundle E with $\cha (E)=\sum_{a}q_a\cha (S^{[a]})^{\ast}$.
(See also references\cite{DD0006,Sug0003A,Hos0007,GJ0010,Tom0010,May0010}
for compact Calabi-Yau cases.)

Our analyses are based on investigation of the geometric properties of
$M$. But we started originally a singular Calabi-Yau manifold in the
Gepner model. In the next section, we will explain relations 
between results in this subsection and those in the Calabi-Yau case.

\subsection{Singular Calabi-Yau manifold}

In this subsection, we study a singular Calabi-Yau $n$-fold $\mfd$
realized as a zero locus in a weighted projective space 
\ba
&&(z_0,z_1,z_2,\cdots ,z_r)\in \Cb P^r(-(r-N),1,1,\cdots ,1)\,,\non
&&\mfd\,;\,p=\mu_0{z}_0^{-u}+\mu_1{z}_1^N+
\mu_2{z}_2^N+\cdots +\mu_r{z}_r^N+\mu_P
\cdot \prod^{r}_{i=0}{z}_i\,,\label{cy}\\
&&u=\frac{N}{r-N}\,,\,\,\,
r=n+1\,,\non
&&\td (T\mfd)=\left(\frac{H}{1-e^{-H}}\right)^r
\left(\frac{1-e^{-NH}}{NH}\right)
\left(\frac{1-e^{-(r-N)H}}{(r-N)H}\right)\,.\nom
\ea
In the context of local mirror symmetries, this singular Calabi-Yau 
manifold can be interpreted as a total space of a bundle 
$\co (-N)\otimes \co (-(r-N))$ over $\X=\Cb P^{r-1}$ and its toric data
are encoded in a vector $\ell^{(0)} $
\ba
\ell^{(0)} =(-N,-(r-N),1,1,\cdots ,1)
\,.\label{data}
\ea
The last term in Eq.(\ref{cy}) induces a deformation of the complex
structure of the manifold and its moduli space is described by 
a set of periods $\Pi_\rho$'s. 
The $\Pi_{\rho}$'s are functions of a variable $w$;
\ba
&&{w}=
(-1)^r\cdot 
\frac{\dis \mu_0^{r-N}\cdot \mu_P^N}{\dis \prod^r_{i=1}\mu_i}\,,\nom
\ea
and we can evaluate their behaviors near $w\sim 0$
\ba
&&\K:=G.C.M.\{N,r-N\}\,,\non
&&\Pi_\rho \sim w^{\rho} \qquad (w\sim 0)\,,\non
&&\qquad \qquad \mbox{for}\,\,\, 
\rho =0,\frac{m_1}{N},\frac{m_2}{r-N}\,\,\mbox{and}\,\,\,
\rho \neq \frac{m}{\K}\,,\non
&&\qquad\qquad (1\leq m_1\leq N-1\,;\,
1\leq m_2\leq r-N-1)\,,\non
&&\Pi_\rho \sim w^{\rho}\log w\,,\,\,\,w^{\rho} \qquad (w\sim 0)\,,\non
&&\qquad \qquad \mbox{for}\,\,\, 
\rho = \frac{m}{\K}\,,\qquad (1\leq m\leq \K-1 )\,.\nom
\ea
The total number of periods is $(r-1)$ and 
$(\K-1)$ solutions have logarithmic behaviors near $w\sim 0$.
Precise formulae of these periods are summarized in the appendix B.

In the previous sections, we investigate properties of this model 
at an orbifold point $\mu_P\sim 0$ based on the Gepner model. 
When we look at the set of periods, some set of these $\Pi_{\rho}$
(in the Cases I,II in the appendix B) are combined 
into a basis of a $\Zb_N$ symmetry at the orbifold point $w=0$
\ba
&&\Pi_{m/N}(e^{2\pi i}w)=e^{2\pi i \frac{m}{N}}\Pi_{m/N} (w)\,,\non
&&\rho =\frac{m}{N}\qquad (m=0,1,2,\cdots ,N-1)\,.\nom
\ea
The $\Zb_N$ action is diagonalized on this set and 
this is an appropriate basis to describe structures near 
the orbifold point in the moduli space.
It corresponds to $N$ elements of the basis in the 
Gepner model we discussed in the section 3.
Now we note that 
the limit $w\rightarrow 0$ is also realized when $\mu_0$ in $\mfd$
tends to zero.
The parameter $\mu_0$ is a
coefficient of the singular part $z_0^{-u}$
and this operation $\mu_0\rightarrow 0$ 
turns out to reduce $\mfd$ to $M$ formally.
That is the reason why the geometric properties of $M$ appear at the 
orbifold point.

Next we continue 
these $\Pi_{\rho}$ $(\rho =\frac{m}{N}\,;\,m=1,2,\cdots ,N-1)$ 
analytically in a large radius region 
and investigate their classical parts $\Pi^{cl}_{\rho}$.
Because we want to 
clarify their geometrical properties, it is enough to restrict ourselves
to these geometric parts $\Pi^{cl}_{\rho =m/N}$
\ba
&&\Pi^{cl}_{m}\equiv \Pi^{cl}_{\rho =m/N}:=
\sum^{N-1}_{a=1}e^{-2\pi i\frac{am}{N}}Z(R_{[a]})\,,\non
&&Z(R_{[a]}):=\sum_{n\geq 0}Z(R_{a+Nn})\,,\non
&&Z(R_b):=
\int_M
\left[
\cha (R_{b})^{\ast}
e^{-tH}\cdot 
\sqrt{\hat{A}(H)}\right]\,,\non
&& \hat{A}(H)=\td (TM)\cdot \td (\co (-(r-N)))\,.\nom
\ea
Here the Chern characters of bundles  $R_a$ or equivalence classes 
$R_{[a]}$ appear naturally and 
these are combined into an appropriate basis to 
describe properties in the orbifold point.
Also we show full formulae of the $\Pi_{\rho=m/N}$ in the appendix C.

Our results based on the CFT are completely satisfactory and 
have appropriate 
geometrical interpretations. But the singular Calabi-Yau manifold 
has vanishing cohomology elements indicated in the paper \cite{Yam0007}. 
They are interpreted to be missing objects in the CFT calculation 
as pointed in \cite{Yam0007,Miz0009,NN0010}. 
Our analyses in the section 3 are based on the Gepner model and 
we do not have enough understanding of these vanishing elements.
The study of them might be possible in the language of local mirror
symmetries.  But we do not touch on them and postpone investigations of 
these singular properties in a future work.

\section{Conclusion}

In this paper, we developed a method to construct the boundary states in
terms of the Gepner-like description of a noncompact singular Calabi-Yau
manifold.  We realize the singular Calabi-Yau manifold as a product of
the Liouville part, $S^1$ part and tensor products of minimal models.

First we analyzed boundary conditions that can be imposed on fields in
the Liouville and $S^1$ sectors.  There are consistent A-,B-type
conditions but the Liouville field always must take a Neumann type
condition.  We find that this fact is reduced to the structure of the
manifold $X$. It is represented as a $\Cx$ fibration over $X/\Cx$ and
the $\Cx$ part is extended in the noncompact direction.  Also the model
has a linear dilaton background and the Liouville field is necessarily
extended in the noncompact direction. It is the reason why the boundary
condition of the Liouville field are free (Neumann type) in this $\Cx$
direction and it induces a trivial structure.

In section 3, we construct boundary states in the A-,B-type cases and
calculate the open string Witten indices between the boundary states.
In both cases, they have a trivial factor $\theta_1(\tau)$.  It has its
origin on the trivial structure of the $\Cx$ direction.  Also that is
related with the Neumann type condition we impose on the Liouville
field. In addition to this trivial factor, the indices have nontrivial
factors.  We can express these factors explicitly and analyze their
geometrical properties.  Especially they are related with B-type
D-branes wrapping around the fibered space $(\text{cycle of } X/\Cx)
\times_{\rm f} \Cx$.

In this paper, we only treat the Neumann boundary condition for the
linear dilaton, in other words, D-branes wrapped on noncompact cycles on
the noncompact manifold. It remains an important problem whether we are
able to consider the D-branes wrapped on compact cycles on the
noncompact manifold i.e. vanishing cycles.  To consider the vanishing
cycles, we need to impose some Dirichlet-like boundary condition on the
$\Ncal=2$ Liouville theory in the noncompact direction.  When one
naively imposes a Dirichlet condition on the $\phi$, the boundary
condition of stress tensor $T$ is broken.  It makes the analysis of
$\phi$ difficult and we do not touch on this problem in this paper.

In section 4, we compare the $\text{cycle of } X/\Cx$ and the nontrivial
factor of the open string Witten index of the B-type boundary
states. They have suitable geometrical interpretations and associated
Witten indices coincide with pairings of coherent sheaves of the
manifold.  That is to say, the boundary states constructed here are
identified with the $S^a$'s (or equivalence classes $S^{[a]}$'s).
It is analogous to the compact Calabi-Yau
cases\cite{DD0006,Sug0003A,Hos0007,GJ0010,Tom0010,May0010}.  However for
the compact Calabi-Yau cases, ambient spaces play crucial roles in the
interpretations of the states in the geometrical language. In these
cases, coherent sheaves of the ambient spaces are related with the
boundary states through restrictions on the hypersurfaces.  In contrast,
for our noncompact Calabi-Yau cases, the Ricci positive manifolds $M$ in
the singular CY's essentially encode information on geometrical
properties of the boundary states.  The noncompact direction controlled
by the Liouville field $\phi$ seems to lead a trivial contribution to
our analyses because we always take a Neumann type condition on the
$\phi$.  When we discard this trivial contribution associated with the
$\phi$, the remaining parts could be understood from the geometric data
of the $M$.

Our results based on the CFT are completely satisfactory and have
appropriate geometrical interpretations.  But the singular Calabi-Yau
manifold has vanishing cohomology elements indicated in the paper
\cite{Yam0007}.  They are interpreted to be missing objects in the CFT
calculation as pointed in \cite{Yam0007,Miz0009,NN0010}.  Our analyses
in the section 3 are based on the Gepner model and we do not have enough
understanding of these vanishing elements.  In fact, we calculated the
full open string Witten index between the boundary states and showed it
is actually zero. The reason why the index vanishes seems to be that we
treat the $\Ncal=2$ Liouville theory as a free field theory and set the
``cosmological constant'' $\mu_0$ to $0$. The geometrical meaning of
this is that the singularity is not deformed.  In \cite{ES0011}, it is
proposed that if the Liouville potential term is treated appropriately,
then nonzero intersection numbers could be obtained.  But our analyses
here are based on the CFT calculations at the Gepner point and it seems
that we study models with the ``cosmological constant'' $\mu_0=0$. That
is consistent with the result\cite{ES0011}.  More precise studies of
them might be possible in the language of local mirror symmetries.  But
we do not touch on them and postpone investigations of these singular
properties in a future work.

After we had submitted this article in the hep-th archive, a paper
\cite{ES0011} by T.~Eguchi and Y.~Sugawara appeared which discusses a
subject related to this article.

\subsection*{Acknowledgement}

The authors would like to thank Michihiro Naka, Masatoshi Nozaki, Yuji
Satoh and Yuji Sugawara for useful discussions and comments.  S.Y. would
also like to thank the organizers (T.~Eguchi et al) of the Summer
Institute 2000 at Yamanashi, Japan, 7-21 August, 2000, where a part of
this work is done.

The work of S.Y. is supported in part by the JSPS Research Fellowships for
Young Scientists.
\newpage
\section*{Appendix A. \ \ \ Theta functions and characters}
In this appendix A, we collect several notations and summarize
properties of theta functions.  We use the following notations in this
paper;
\begin{eqnarray*}
 &&\e{x}:=\exp(2\pi i x),\\
 &&\deltam{m}{N}:=
\begin{cases}
 1 & (m\equiv 0 \mod N),\\
 0 & ({\rm others}),
\end{cases}\\
&& \deltam{m,m'}{N}:=\deltam{m-m'}{N},
\end{eqnarray*}
where $m$ and $N$ are integers.  A useful equation is satisfied for
integers $m$ and $N$
\begin{eqnarray*}
 &&\sum_{j\in \Zb_N}\e{\frac{jm}{N}}=N\deltam{m}{N}.
\end{eqnarray*}
A set of SU(2) classical theta functions are defined as
\begin{eqnarray*}
&& \Th_{m,k}(\tau,z)=\sum_{n\in \Zb}q^{k\left(n+\frac{m}{2k}\right)^2}
 y^{k\left(n+\frac{m}{2k}\right)},
\end{eqnarray*}
with $q:=\e{\tau},y:=\e{z}$.  The Jacobi's theta functions are also
defined in our convention
\begin{eqnarray*}
&& \theta_{1}(\tau,z):=i\sum_{n\in \Zb}(-1)^n q^{\left(n-\frac{1}{2}\right)^2}
 y^{\left(n-\frac{1}{2}\right)},
 \theta_{2}(\tau,z):=\sum_{n\in \Zb} q^{\left(n-\frac{1}{2}\right)^2}
 y^{\left(n-\frac{1}{2}\right)},\\
&& \theta_{3}(\tau,z):=\sum_{n\in \Zb} q^{n^2}y^{n},\hspace{2.7cm}
 \theta_{4}(\tau,z):=\sum_{n\in \Zb}(-1)^n q^{n^2}y^{n}.
\end{eqnarray*}
The above two kinds of theta functions are related through a set of
linear transformations
\begin{eqnarray*}
 && 2\Th_{0,2}=\theta_3+\theta_4 ,\ 2\Th_{1,2}=\theta_2+i\theta_1,\ 
 2\Th_{2,2}=\theta_3-\theta_4 ,\ 2\Th_{3,2}=\theta_2-i\theta_1.
\end{eqnarray*}
The Dedekind $\eta$ function is represented as an infinite product
\begin{eqnarray*}
 \eta(\tau):=q^{\frac1{24}}\prod_{n=1}^{\infty}(1-q^n).
\end{eqnarray*}
The character $\chi_s(\tau,z),\ s=0,1,2,3$ of $\widehat{SO(d)}_1$ for
$d/2 \in 2\Zb+1$ can be expressed as
\begin{eqnarray*}
 && \chi_0(\tau,z)=\frac{\theta_3(\tau,z)^{d/2}
+\theta_3(\tau,z)^{d/2}}{2\eta(\tau)^{d/2}},\ \ 
  \chi_1(\tau,z)=\frac{\theta_2(\tau,z)^{d/2}
+(i\theta_1(\tau,z))^{d/2}}{2\eta(\tau)^{d/2}},\\
 && \chi_2(\tau,z)=\frac{\theta_3(\tau,z)^{d/2}
-\theta_3(\tau,z)^{d/2}}{2\eta(\tau)^{d/2}},\ \ 
  \chi_3(\tau,z)=\frac{\theta_2(\tau,z)^{d/2}
-(i\theta_1(\tau,z))^{d/2}}{2\eta(\tau)^{d/2}}.
\end{eqnarray*}
Next we introduce a character $\chi_m^{\ell,s}(\tau,z)$ of a Verma
module $(\ell,m,s)$ in the level $(N-2)$ minimal model.  This function
has equivalence relations with respect to its indices
\begin{eqnarray*}
 \chi_m^{\ell,s}=\chi_{m+2N}^{\ell,s}=\chi_m^{\ell,s+4}
=\chi_{m+N}^{N-2-\ell,s+2}.
\end{eqnarray*}
One can see an explicit form of this $\chi_m^{\ell,s}(\tau,z)$ in the
paper\cite{Gep88}.

 Next we shall look at modular properties of these functions.  They
behave under the T transformation $\tau\rightarrow \tau +1$
\begin{eqnarray*}
 && \Th_{m,k}(\tau+1,z)=\e{\frac{m^2}{4k}}\Th_{m,k}(\tau,z),\\
 && \theta_1(\tau+1,z)=\e{\tfrac18}\theta_1(\tau,z), 
 \qquad \theta_2(\tau+1,z)=\e{\tfrac18}\theta_2(\tau,z),\\
 && \theta_3(\tau+1,z)=\theta_4(\tau,z), 
 \qquad \theta_4(\tau+1,z)=\theta_3(\tau,z),\\
 && \eta(\tau+1)=\e{1/24}\eta(\tau),\\
 && \chi_s(\tau+1,z)=\e{\frac{s^2}{8}-\frac{d}{48}}\chi_s(\tau,z),\\
 && \chi_m^{\ell,s}(\tau+1,z)=\e{\frac{\ell(\ell+2)}{4N}-\frac{m^2}{4N}
+\frac{s^2}{8}-\frac{N-2}{8N}}\chi_m^{\ell,s}(\tau,z).
\end{eqnarray*}
For the S transformation $\tau\rightarrow -1/\tau$, they have following
modular properties
\begin{eqnarray*}
 &&\Th_{m,k}(-1/\tau,z/\tau)=
\sqrt{-i\tau}\e{\frac{k}{4}\frac{z^2}{\tau}}
\sum_{m'\in \Zb_{2k}}\frac{1}{\sqrt{2k}}\e{-\frac{mm'}{2k}}
\Th_{m',k}(\tau,z),\\
 &&\theta_1(-1/\tau,z/\tau)=-i\sqrt{-i\tau}\e{\frac12 \frac{z^2}{\tau}}
\theta_1(\tau,z),
\qquad \theta_2(-1/\tau,z/\tau)=\sqrt{-i\tau}\e{\frac12 \frac{z^2}{\tau}}
\theta_4(\tau,z),\\
 &&\theta_3(-1/\tau,z/\tau)=\sqrt{-i\tau}\e{\frac12 \frac{z^2}{\tau}}
\theta_3(\tau,z),
\qquad \theta_4(-1/\tau,z/\tau)=\sqrt{-i\tau}\e{\frac12 \frac{z^2}{\tau}}
\theta_2(\tau,z),\\
&& \eta(-1/\tau)=\sqrt{-i\tau}\eta(\tau),\\
&&\chi_s(-1/\tau,z/\tau)=\e{\frac{d}{4}\frac{z^2}{\tau}}
\sum_{s'=0}^{3}\frac 12 \e{-\frac d2 \frac{ss'}{4}}\chi_{s'}(\tau,z),\\
 && \chi_m^{\ell,s}(-1/\tau,z/\tau)=
\e{\frac{N-2}{2N}\frac{z^2}{\tau}}
\frac{1}{\sqrt{8N}} \sum_{\ell,m,s}^{\rm even}
A_{\ell\ell'}\e{-\frac{ss'}{4}+\frac{mm'}{2N}}
\chi_{m'}^{\ell',s'}(\tau,z),\\
&&A_{\ell\ell'}=\sqrt{\frac2N}\sin\left[\pi\frac{(\ell+1)(\ell'+1)}{N}\right].
\end{eqnarray*}
Here the symbol $\sum_{\ell,m,s}^{\rm even}$ means sums  
for $(\ell,m,s)$ under conditions $\ell+m+s \equiv 0\mod 2$.
Also we use the notation $f(\tau)$ for a function $f(\tau,z)$
of $\tau,z$ with substituting $z=0$.

The SU(2) fusion coefficients $N_{\ell_1\ell_2}^{\ell_3}$ for
$\ell_1,\ell_2,\ell_3=0,1, \dots,N-2$
are calculated in the following form 
\begin{align*}
  N_{\ell_1\ell_2}^{\ell_3}=
\begin{cases}
 1  & (|\ell_1-\ell_2|\le \ell_3 \le \min\{\ell_1+\ell_2,2N-4-\ell_1-\ell_2\})\\
 0  & (\text{others})
\end{cases}.
\end{align*}
We can extend this definition satisfied for all integer
$\ell_1,\ell_2,\ell_3$ by using relations $
N_{\ell_1\ell_2}^{\ell_3}=N_{\ell_1\ell_2}^{-\ell_3-2}
=N_{\ell_1\ell_2}^{\ell_3+2N}.  $ Then, the Verlinde formula
\begin{align*}
 N_{\ell_1\ell_2}^{\ell_3}=\sum_{\ell=0}^{N-2}
\frac{A_{\ell\ell_1}A_{\ell\ell_2}A_{\ell\ell_3}}{A_{0\ell}}
\end{align*}
is satisfied.

\newpage
\section*{Appendix B. \ \ \ Periods near the orbifold point}

We will write down periods $\Pi_{\rho}$ for the Calabi-Yau manifold
$\mfd$ in the region near the orbifold point.  These solutions are
labelled by a variable $\rho$;
\begin{enumerate}
\item[] \underline{Case I} ; $\rho =0$
\begin{eqnarray}
\Pi_{\rho} \equiv 1\,\nom
\end{eqnarray}
\item[] \underline{Case II} ; $\rho =\frac{s+1}{N}$ 
$(s=0,1,2,\cdots ,N-2)$ 
\begin{eqnarray}
&&{\Pi_{\rho}}=
\sum_{m\geq 0}
\frac{\dis \left[\gam \left(m+\frac{s+1}{N}\right)\right]^{n+2}
\times {w}^{m+\frac{s+1}{N}}}
{\dis \gam (Nm+s+1)\gam \left(m+1+\frac{s+1}{N}\right)
\gam \left((r-N)m+\frac{r-N}{N}(s+1)\right)}
\,,\nom
\end{eqnarray}
\item[] \underline{Case III} ; $\rho 
=\frac{s+1}{r-N}$ $(s=0,1,2,\cdots ,r-N-2)$ and $\rho \neq \frac{m}{\K}$
\begin{eqnarray}
&&{\Pi_{\rho}}=
\sum_{m\geq 0}
\frac{\dis \left[\gam \left(m+\frac{s+1}{r-N}\right)\right]^{n+2}
\times {w}^{m+\frac{s+1}{r-N}}}
{\dis \gam ((r-N)m+s+1)\gam \left(m+1+\frac{s+1}{r-N}\right)
\gam \left(Nm+\frac{N}{r-N}(s+1)\right)}
\,,\nom
\end{eqnarray}
\item[]
\underline{Case IV} ; $\rho =\frac{s+1}{\K}$ $(s=0,1,2,\cdots ,\K-2)$ 
\begin{eqnarray}
&&{\Pi_{\rho}}=
\log {w}\times
\sum_{m\geq 0}
\frac{\dis \left[\gam \left(m+\frac{s+1}{\K}\right)\right]^{n+2}
\times {w}^{m+\frac{s+1}{\K}}}
{ \gam (\frac{r-N}{\K}(s+1)+(r-N)m)\gam \left(m+1+\frac{s+1}{\K}\right)
\gam \left(Nm+\frac{N}{\K}(s+1)\right)}\non
&&\qquad +
\sum_{m\geq 0}
\frac{\dis \left[\gam \left(m+\frac{s+1}{\K}\right)\right]^{n+2}
\times {w}^{m+\frac{s+1}{\K}}}
{ \gam (\frac{r-N}{\K}(s+1)+(r-N)m)\gam \left(m+1+\frac{s+1}{\K}\right)
\gam \left(Nm+\frac{N}{\K}(s+1)\right)}\non
&&\times \Bigl\{
(n+2)\Psi\left(\frac{s+1}{\K}+m\right)-N\cdot 
\Psi\left(\frac{N}{\K}(s+1)+Nm\right)\non
&&\qquad 
-(r-N)\Psi \left(\frac{r-N}{\K}(s+1)+(r-N)m\right)
-\Psi\left(1+m+\frac{s+1}{\K}\right)
\Bigr\}
\,.\nom
\end{eqnarray}
\end{enumerate}
We use solutions in the Cases I, II to discuss $\Zb_N$ properties of the
$M$. The solutions in the Case IV have logarithmic behaviors near the
orbifold point.

\newpage
\section*{Appendix C. \ \ \ Periods in the large volume region}

We summarize concrete formulae of the $\Pi_{\rho}$'s for $\rho=m/N$
$(m=1,2,\cdots ,N-1)$ in the large volume region
\begin{eqnarray}
&&\Pi_{\rho}=(2\pi i)^{r-1}
\times {(r-N)}\sum^{N-1}_{a=1}e^{-2\pi i \rho a}
\non
&&\qquad \times \sum_{n\geq 0}
\int_M
\left[
\cha (R_{a+Nn})^{\ast}
e^{-tH}\cdot 
\sqrt{\hat{A}(H)}\cdot \sqrt{\hat{K}(-H)}
\times \exp \left(
\sum_{\ell \geq 2}(-H)^{\ell}x_{\ell}
\right)
\right]\,,\non
&&t=\frac{1}{2\pi i}\left[
\log {z}+\sum_{m\geq 1}a(m){z}^m
\right]\,,\qquad z=w^{-1}\,,\non
&&\hat{A}(H)=
\left(\frac{\dis\frac{H}{2}}{\dis \sinh \frac{H}{2}}\right)^r\cdot 
\left(\frac{\dis \sinh \frac{NH}{2}}{\dis \frac{NH}{2}}\right)
\cdot 
\left(\frac{\dis \sinh \frac{(r-N)H}{2}}{\dis
      \frac{(r-N)H}{2}}\right)\,,\non
&&\hat{K}(-H)=
\exp\left[
\sum^{\infty}_{\ell =1}
\frac{\zeta (2\ell +1)}{2\ell +1}\cdot 
\left(\frac{H}{2\pi i}\right)^{2\ell +1}
\times \{N^{2\ell +1}+(r-N)^{2\ell +1}-r\}
\right]\,,\non
&&x_{\ell}=
\frac{1}{(2\pi i)^{\ell}}\cdot
\frac{1}{\ell !}
\del_{v}^{\ell}\log
\left\{
1+v \sum_{m\geq 1}\frac{a(m+v)}{\tilde{a}(v)}{z}^m
\right\}_{v =0}\,,\non
&& a(m+v)=
\frac{\dis \gam (1+N(m+v))\gam (1+(r-N)(m+v))\gam (m+v)}
{[\gam (1+m+v)]^{n+2}}\,,\non
&&\tilde{a}(v)=
\frac{\dis \gam (1+Nv)\gam (1+(r-N)v)}{[\gam (1+v)]^r}\,.\nom
\end{eqnarray}
These encode information on bundles $R_a$ and their properties are 
discussed in section 4.

\newpage
\providecommand{\href}[2]{#2}\begingroup\raggedright\endgroup

\end{document}